%% file: main.tex
\documentclass[10pt,journal]{IEEEtran}
\usepackage[T1]{fontenc}
\usepackage{amsmath,amssymb,booktabs,tabularx,graphicx,microtype}
\usepackage[hidelinks]{hyperref}
\newcommand{\E}{\mathbb{E}}
\newcommand{\Rea}{\operatorname{Re}}
\newcommand{\tr}{\operatorname{tr}}

\newcommand{\norm}[1]{\left\lVert#1\right\rVert}

\title{Negative Results for Reciprocity-Based Over-the-Air Split Learning Under Channel Aging: Stability, Alignment, and Geometry}
\author{\IEEEauthorblockN{Tianhao Guo, Xianzhong Li, and Hao Luo}\\
\IEEEauthorblockA{College of Physics and Electronic Engineering, Shanxi University, Taiyuan 030006, China. Email: tianhao\_guo@sxu.edu.cn}\\
\IEEEauthorblockA{Corresponding author: Tianhao Guo}}
\date{}
\begin{document}
\maketitle
\begin{abstract}
In reciprocity-based over-the-air split learning, channel aging rotates and attenuates the derivative signal returned to the transmitter. Seven case studies combine numerical forecasts, mechanism checks, and competing predictions. Near the correlation zero, an archived SGD comparison finds truncated inversion 0.000813 lower in cross-entropy than polar correction, with a paired 95\% interval for the difference from $-0.001853$ to 0.000060, and a retained-mode quadratic proxy slightly favors truncation. Selecting the return port with the largest received power falls short of the predicted five-point alignment advantage over random selection at both tested spacings, gaining at most 0.97 points, while the local inverse relation between amplitude and phase gradient is supported. Seven specified interference-induced accuracy-collapse predictions are rejected, and independent zero-mean interference preserves the conditional mean under a fixed linear chain. Per-subcarrier correction and spherical-wavefront compensation do not deliver their predicted accuracy gains. Focusing leaves both methods near chance in a single-mode model and adds no observed gain in a model with a line-of-sight component plus scattering. A successful port assignment under nonsymmetric motion refutes an exact-symmetry requirement. The competing positioning-error prediction is supported. Gaussian rail jitter with a standard deviation of 0.383 wavelengths yields high terminal test accuracy, whereas a fixed bias of the same magnitude yields near-chance accuracy. These cases distinguish valid local physical relations from unsupported training forecasts through task support, noise injection, and update moments.
\end{abstract}

\section{Setting and evidence protocol}
Over-the-air split learning computes part of the backward pass in the radio channel: the receiver sends the derivative of its loss back through the reverse link, and time-division-duplex reciprocity lets the channel apply the required adjoint. When the terminal moves between the forward and reverse transmissions, channel aging rotates and attenuates this returned derivative. Several remedies and explanations suggest themselves from communication practice: avoid matrix inversion, select the strongest port, suppress interference, correct each subcarrier, or exploit array symmetry. The empirical comparisons use predictions and falsification lines fixed before each run. We also report analytic counterexamples. This paper separates rejected performance forecasts, mechanism counterexamples, and supported competing predictions. Companion studies report the designs that survived for this setting: movable antennas that keep pace with the user~\cite{jsac}, waiting out the dead zones of a single aging antenna~\cite{wcl}, and statistical repair of the return at a fixed array~\cite{tcom}. The present paper records the forecasts that did not.

Write the forward and return channel matrices as $H_f$ and $H_b$. For a linear minimum-mean-square predictor, a useful representation is~\cite{aging}
\begin{equation}
 H_b=AH_f+\Delta H,\qquad A=C_{10}C_{00}^{\dagger},
 \label{eq:channel}
\end{equation}
where $C_{00}=\E[H_fH_f^H]$ and $C_{10}=\E[H_bH_f^H]$, with the predictor restricted to the signal support when necessary. If the transmitted derivative seed is $M\delta$, a representative received derivative is $H_b^HM\delta+n$. This distinguishes innovation multiplied by $M\delta$ from receiver noise $n$ injected after that multiplication. The distinction determines whether an inverse amplifies a given noise term.

The quantitative training predictions are working hypotheses proposed during this study. The cited classical works supply the physical or algebraic premises from which those predictions were extrapolated.
The training comparisons use eight seed-level repetitions per reported configuration. Accuracy is the terminal test accuracy, cross-entropy (CE) uses natural logarithms, and accuracy differences are percentage points (pp). Confidence intervals reproduce the archived seed-level analyses. Related methods share seeds. Subcarriers, test examples, and time steps are not additional independent training repetitions. Test A also contains an analytic quadratic proxy, and Test B includes channel-only measurements.

\subsection{Tasks and training configurations}
All classifier experiments use synthetic inputs $x\in\{-1,+1\}^3$ with independent equiprobable coordinates and binary labels $\ell=\mathbf{1}[x_0x_1>0]$. A complex linear encoder $W$ feeds the forward channel. The receiver applies $\tanh(|y|^2)$ componentwise and a real linear logistic head. Table~\ref{tab:tasks} gives the encoder dimensions and number of receiver features, including separate models on each OFDM tone. Every classifier configuration uses batch size $32$, eight evaluation seeds, and $8192$ test examples (per tone for wideband tests). Adam uses fixed encoder/receiver learning rates $0.025/0.06$, $(\beta_1,\beta_2)=(0.9,0.99)$ and $\epsilon=10^{-8}$. The encoder is projected onto $\norm{W}_F\leq1/\sqrt{3}$, receiver coordinates onto $[-4,4]$, with receiver weight regularization $0.001$. The SGD exceptions and training budgets are explicit in the table.

\begin{table*}[t]
\caption{Classifier configurations from the archived protocols and implementation. All rows use the binary task, batch size, evaluation seeds and test size defined in Section~I. Dimensions are complex encoder entries / real receiver features. ``Per tone'' means independently trained parameters. Noise entries are proper-complex variances per coordinate before gradient-gain recovery. Interference and phase distortion are additional to the return thermal noise. A dash in the CSI column means no CSI arm is used in that comparison.}
\label{tab:tasks}\centering\footnotesize
\renewcommand{\arraystretch}{1.12}
\begin{tabularx}{\textwidth}{@{}l l r r l X@{}}\toprule
Test & Encoder / features & Examples / steps & CSI examples / steps & Optimizer & Forward / return variance; additional distortion \\\midrule
A (SGD) & $2\!\times\!3$ / $3$ & $3840/120$ & --- & SGD$^{a}$ & $0.01/0.0009$ \\
B (replay) & $4\!\times\!3$ / $1$ & $7680/240$ & $7680/240$ & Adam & $0.01/0.0009$; eight candidate return ports \\
C1 & $4\!\times\!3$ / $1$ & $3840/120$ & $2666/84$ & Adam & $0.01/0.0009$; interferer/target power ratio $10$ \\
C2 & $4\!\times\!3$ / $1$ & $3840/120$ & $2666/84$ & Adam & $0.01/0.0009$; interferer/target power ratio $10$ \\
C3 & $4\!\times\!3$ / $1$ & $3840/120$ & $2666/84$ & Adam & $0.01/0.0009$; total interference/target ratio $20$ \\
C4 & $4\!\times\!3$ / $1$ & $3840/120$ & $2666/84$ & Adam & $0.01/0.0009$; self-interference $20$ dB \\
C5 & $2\!\times\!3$ / $1$ (single AP) & $3840/120$ & $2666/84$ & Adam & $0.01/0.0009$; full-network encoder $8\!\times\!3$ \\
C6 & $4\!\times\!3$ / $1$ & $3840/120$ & $2666/84$ & Adam & $0.01/0.0009$; two self-interferers, total $20$ dB \\
C7 & $8\!\times\!3$ / $1$ & $3840/120$ & $2666/84$ & Adam & $0.01/0.0009$; AP phases $U[-\pi/3,\pi/3]$, $4$-bit fronthaul \\
C8 & $4\!\times\!3$ / $1$ & $3840/120$ & $2666/84$ & Adam & $0.01$ / $10^{-s/10}$, $s=0,10,20$ dB; multiuser arms: $0.0009$ plus total ISR $0,10,20$ dB \\
C9 & $4\!\times\!3$ / $1$ & $3840/120$ & $3840/120$ & SGD/Adam$^{b}$ & $0.01/0.0009$; total SIR $0$ dB \\
C10 & $4\!\times\!3$ / $1$ & $3840/120$ & $3840/120$ & Adam & $0.01/0.0009$; SIR $0$ dB \\
D1 & $2\!\times\!3$ / $3$, per tone & $3840/120$ & $2666/84$ & Adam & $0.01/0.0009$; eight OFDM tones \\
D2 & $2\!\times\!3$ / $3$, per tone & $3840/120$ & $2666/84$ & Adam & $0.01/0.0009$; eight OFDM tones \\
D3 & $2\!\times\!3$ / $4$, per tone & $3840/120$ & $2666/84$ & Adam & $0.01/0.0009$; eight OFDM tones \\
E1 & $2\!\times\!3$ / $8$ & $3680/115$ & $2666/84$ & Adam & $0.01/0.0009$ \\
E2 & $2\!\times\!3$ / $8$ & $3680/115$ & $2666/84$ & Adam & $0.01/0.0009$ \\
F1 & $2\!\times\!3$ / $7$ & $3840/120$ & $2666/84$ & Adam & $0.01/0.0009$; return recovery factor $8\sqrt{7}$ \\
F2 & $2\!\times\!3$ / $7$ or $9$ & $3840/120$ & $3840/120$ & Adam & $0.01/0.0009$; return recovery factor $8\sqrt{N_r}$ \\
G1 & $2\!\times\!3$ / $4$ & $3840/120$ & $2666/84$ & Adam & $0.01/0.0009$; return recovery factor $16$ \\
G2 & $2\!\times\!3$ / $4$ & $3840/120$ & $3840/120$ & Adam & $0.01/0.0009$; return recovery factor $16$ \\\bottomrule
\end{tabularx}
\begin{minipage}{\textwidth}\footnotesize
$^{a}$No momentum. Encoder and receiver share the method-specific rate selected on three separate tuning seeds from $\{0.01,0.03,0.1,0.3,1,3,10\}$. The archived choice is $0.3$ for both methods.
$^{b}$Both optimizers use encoder/receiver rates $0.025/0.06$. SGD has no momentum. CSI in C9 uses Adam.
CSI arms bypass return thermal noise. Their forward noise is unchanged. Other physical arms use per-example derivative-norm recovery except the fixed factors shown for F and G. F1's polar reference uses $3712$ examples and $116$ steps. Each D-row budget is per tone. C5 selects the best of four separately trained single-AP models using terminal test accuracy. C7's additional single-AP reference uses $2\!\times\!3$ encoders. All learning rates are constant within a run.
\end{minipage}
\end{table*}

The quadratic part of A instead minimizes $f(x)=\norm{x}^2/2$ on two retained or all three channel modes, with deterministic $H_f=U_r$, no forward noise, independent innovation columns of covariance $Q/r$, and downstream noise $\mathcal{CN}(0,0.03^2I)$. Its stochastic linear update uses step size $0.3$ and eight seeds, with one noise realization per step and no data minibatch or classifier. Stable cases run $20000$ steps, discarding $5000$. Unstable cases run $1000$ steps and use the final $500$ only as a finite-horizon diagnostic. B's channel-only measurements use eight seeds with $8192$ independent fields per seed, four transmit components and eight candidate return ports. The scalar phase-gradient statistic uses one component. These noiseless field measurements have no class labels, network, batch size, optimizer or training updates. B's replay reads the earlier trained models in the table.

Across the seven cases, the evidence has distinct roles. The port-selection, interference-collapse, wideband-gain, and focusing-gain forecasts cross their numerical rejection lines. Nonsymmetric port assignment supplies a counterexample to a necessity claim. The local amplitude--phase relation and the competing bias-versus-jitter prediction receive support. The quadratic model tests a mechanism through its second-moment recursion, while the nearby SGD comparison estimates a CE difference whose interval spans zero. Appendix~\ref{app:methods} specifies the operators, channel laws, and calibration used in these comparisons.

\section{Unitary Polar Correction Versus Truncated Inversion}
The hypothesis that polar correction should outperform truncated inversion near the correlation zero was proposed in this study, motivated by classical zero-forcing noise amplification~\cite[Sec.~8.3.1]{tse} and the norm-preserving polar approximation~\cite{higham}. An inverse amplifies upstream noise along a weak singular direction, whereas the unitary polar factor preserves norm.

The archived forecast for a three-element half-wavelength array at displacement $0.383\lambda$ required polar CE $\leq0.08$, truncated-inverse CE $\geq0.28$, and a gap exceeding $0.20$, with step size $0.3$ and noise standard deviation $0.03$. Its rejection rule was a same-setting steady-state inverse win, or persistent negative empirical inner products causing early polar divergence. The precise neural-network comparison in that rule remains untested: Test A supplies a quadratic mechanism check and a nearby archived SGD comparison.

The weakest singular value is $0.00915798$. On the two retained task modes, the exact stationary squared-gradient difference, TSVD minus polar, is $-4.03688\times10^{-7}$ (Table~\ref{tab:a}). The paired simulation difference is $-3.66360\times10^{-7}$ with a seed-level $t$ interval $[-4.32692,-3.00029]\times10^{-7}$. In the nearby SGD archive, inverse and polar CE are $0.24962161$ and $0.25043496$: difference $-0.00081335$, with a paired 95\% bootstrap interval $[-0.001853,0.000060]$. The quadratic stationary energy cannot be converted into a CE forecast.

\begin{table}[t]
\caption{Test A: archived quadratic second moments. $r_2$ is the spectral radius of the second-moment operator. Finite stationary values are $\E\norm{\nabla f}^2\times10^4$. A dash denotes no finite steady state.}
\label{tab:a}\centering\footnotesize
\begin{tabularx}{\columnwidth}{@{}Xlrr@{}}\toprule
Task & Method & $r_2$ & Energy \\\midrule
\input{figures/table_a.tex} 
\bottomrule\end{tabularx}
\end{table}

Hard truncation sets the inverse gain of the deleted mode to zero. It does not replace every singular value by the truncation threshold. If the task uses only retained modes, deleting a weak mode need not impose the proposed noise penalty. Conversely, retaining that mode with a unitary correction does not guarantee mean-square stability.
For the archived proxy, let $f(x)=\norm{x}^2/2$, $H_f=U_r$, and $\delta=U_rx$, where $U_r$ contains retained left singular vectors of $A^H$. Within this proxy, each step draws fresh, zero-mean innovation columns with covariance $Q/r$ and downstream noise with covariance $\sigma_w^2 I$, independently of the past and of each other. These conditional moment assumptions are specific to the proxy. Define
\begin{align}
 B&=U_r^HA^HMU_r, & F&=I-\eta B,\nonumber\\
 R&=U_r^HM^H(Q/r)MU_r. &&
\end{align}
The raw second moment $S_k=\E[x_kx_k^H]$ obeys
\begin{equation}
 S_{k+1}=FS_kF^H+\eta^2\tr(RS_k)I+\eta^2\sigma_w^2I.
 \label{eq:moments}
\end{equation}
This is a linear recursion on matrices~\cite{dieuleveut}. Its lifted spectral radius, rather than norm preservation of $M$, controls a finite second-moment steady state. Activating all three modes makes that radius exceed one for polar correction with innovation. TSVD leaves an undamped noisy direction.

In this fixed-frame quadratic model and the nearby SGD archive, comparisons must specify the task support, noise injection point, and stability domain.

\section{Port Selection by Received Power and Cross-Exchange Phase Alignment}
The hypothesis that the brightest return port should preserve cross-exchange phase alignment was proposed in this study by extrapolating the local amplitude--phase relation in \emph{Dislocations in Wave Trains}~\cite{nyeberry}, which identifies amplitude zeros where phase is undefined.

Test B fixes the forward port and selects only the return port from eight candidates using total power across four transmit-channel components. At spacings $0.2\lambda$ and $0.5\lambda$, the prediction required a positive-alignment probability close to geometric nearest-position selection and an advantage over random selection. The rejection line was a max-power minus random advantage below $5$ pp. A separate local prediction required the bright-to-dark phase-gradient ratio and its interval upper bound to be below $0.8$, with decreasing bin means. A ratio above $0.8$ would reject it.

\begin{figure*}[t]
\centering\includegraphics[width=.93\textwidth]{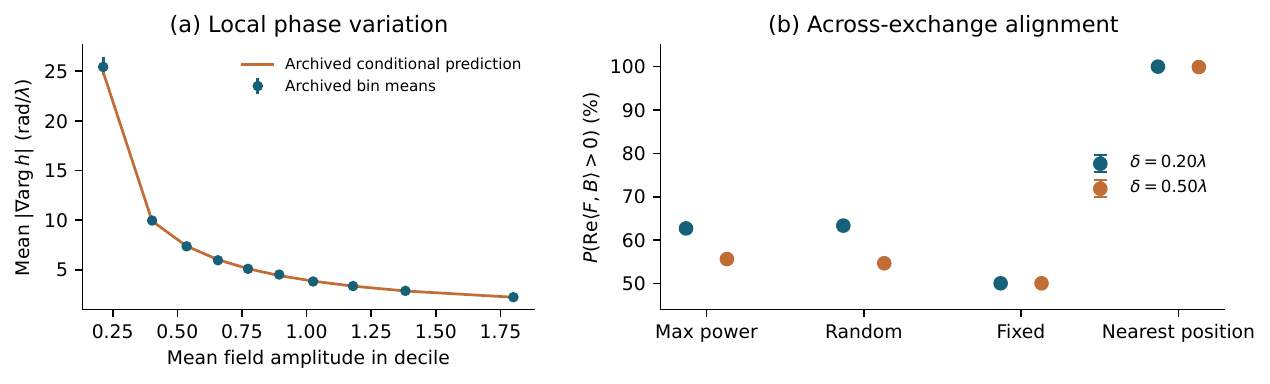}
\caption{Test B separates a valid local relation from an invalid cross-exchange inference. Left: archived amplitude-bin means and the conditional prediction. Right: positive real channel alignment for return-port rules with a common fixed forward port. Error bars reproduce seed-bootstrap 95\% intervals. A sampled probability of one does not establish a population guarantee.}
\label{fig:b}
\end{figure*}

The local ratio is $0.08694$, with interval $[0.08320,0.09042]$. However, max-power minus random alignment is $-0.647$ pp at the finer spacing and $+0.970$ pp at the coarser spacing. Geometric selection reaches sampled probabilities of $100.000\%$ and $99.899\%$ (Fig.~\ref{fig:b}). Both global comparisons cross the rejection line while the local relation holds. A separate replay of the original training channels pairs $89.169\%$ accuracy with only $63.641\%$ positive channel alignment and mean cosine $0.29121$.

For a two-dimensional, zero-mean, unit-variance proper Gaussian field with monochromatic isotropic covariance $\E[h(r)h(r')^*]=J_0(2\pi\norm{r-r'})$, and for $a>0$, the archived calculation gives
\begin{equation}
 \E[\norm{\nabla\arg h}\mid |h|=a]
 =\frac{\pi\sqrt{\pi/2}}{a}
 \label{eq:phase}
\end{equation}
when distance is measured in wavelengths. It neither bounds phase accumulated over a finite displacement nor couples the maximizer of return power to the earlier forward location. Moreover, a large vector-channel norm does not require every scalar component to be bright. The measured probability uses a positive real normalized channel inner product.

The training replay rejects near-perfect per-exchange alignment as a necessary explanation of its accuracy. Conditional gradient alignment, second moments, and step sizes enter standard descent analyses~\cite{bottou}. An unconditional channel cosine does not supply them.

For this return-only selector, local brightness does not explain persistent alignment.

\section{Conditional-Mean Invariance Under Unsuppressed Interference}
The seven collapse predictions below were working hypotheses proposed in this study, extrapolating the detection penalty of interference in classical multiuser communication~\cite[Ch.~8]{tse} to training under a distorted backward signal.

Tests C1--C7 freeze one collapse prediction for each of the seven stress cases in Table~\ref{tab:c}. The last column gives the accuracy that rejects it. C1--C4 test the unsuppressed corrected return, C5 the best single access point (AP), C6 the projected return with two self-interferers, and C7 centralized correction with AP phase jitter. Companion protocols also predict large recovery margins from additional processing. Tests C8--C10 subsequently examine whether gradient signal-to-noise ratio (GSNR), optimizer scaling, or mean-gradient cosine can explain method rankings.

\begin{table}[t]
\caption{Tests C1--C7: selected collapse predictions all cross their own rejection lines. Accuracies and intervals are percentages.}
\label{tab:c}\centering\scriptsize
\begin{tabularx}{\columnwidth}{@{}Xrrr@{}}\toprule
Stress case & Accuracy & 95\% interval & Reject \\\midrule
\input{figures/table_c.tex} 
\bottomrule\end{tabularx}
\end{table}

A fixed-chain diagnostic finds relative changes in the complex gradient projection coefficient of $0.0382\%$ for white noise and $0.8835\%$ for multiuser interference over the archived power comparison, below its $3\%$ support target. The rejection rule is a relative deviation above $10\%$ or a sign reversal of the mean cosine. Deviations between the two thresholds are inconclusive. Yet complete-vector GSNR has accuracy rank correlation $0.3466$, below its $0.80$ line. The alternative scalar proxy reaches $0.8556$, showing that the two definitions are not interchangeable.
ZF loses $8.0261$ pp under SGD and $7.2891$ pp under Adam relative to their respective single-user controls. The loss ratio $1.1011$ is below the frozen rejection line of $2$ for the optimizer-amplification explanation. In a separate ideal-operator direction test, the mean-cosine/accuracy rank correlation is $0.6412$, below its $0.70$ line. These observations leave the historical ZF loss without a single established explanation.

For a fixed linear correction chain $T$, an additive term $\xi$ with $\E[\xi\mid\mathcal F]=0$ satisfies
\begin{equation}
 \E[T(s+\xi)\mid\mathcal F]=T\E[s\mid\mathcal F].
 \label{eq:additive}
\end{equation}
Here $\mathcal F$ fixes the chain and relevant task state. Independent zero-mean contamination can increase variance without creating a systematic reverse direction. This explains why interference power alone is insufficient to predict collapse. The AP tests have another structure: deterministic per-link phases can be corrected locally, and the tested random phase law retains a positive mean complex factor.

An orthogonal ZF projection is contractive before any target-gain inversion. It can remove target energy even while reducing absolute noise. Adam's gradient-rescaling property~\cite{adam} does not restore discarded components, and a mean cosine or a scalar projection discards information about gain, covariance, and task geometry. No one of these metrics determines the full update distribution.

For the frozen linear, independent-message and AP proxies, seven specified collapses fail.

\section{Per-Subcarrier Correction in Wideband Aging Channels}
The hypothesis that wideband aging requires per-subcarrier correction was proposed in this study by extrapolating the subcarrier-dependent instantaneous gains of classical OFDM~\cite[Sec.~3.4]{tse} to the statistical forward--return operator.

Tests D1--D3 predict failure of a global average polar correction and a large advantage from per-subcarrier correction. The relative-accuracy rejection lines are gains below $10$, $6$, and $4$ pp, respectively. D1 uses independent taps with common spatial statistics. D2 assigns distinct angles to taps. D3 additionally makes array and motion phases frequency dependent, at fractional bandwidth $0.20$.

All three relative advantages cross their rejection lines (Table~\ref{tab:d}). In D3 the gain is $0.008202$ pp, with paired interval $[-0.005722,0.020981]$ pp. Its CE advantage is $-0.000014820$, also with an interval spanning zero.

\begin{table}[t]
\caption{Test D: terminal accuracy (\%) and local-minus-global gain (pp). Global denotes a shared statistical polar correction.}
\label{tab:d}\centering\footnotesize
\begin{tabularx}{\columnwidth}{@{}Xrrr@{}}\toprule
Channel model & Global & Local & Gain \\\midrule
\input{figures/table_d.tex} 
\bottomrule\end{tabularx}
\end{table}

Instantaneous frequency selectivity does not imply frequency dependence of $A_k=C_{10,k}C_{00,k}^{\dagger}$. For independent taps, the delay phase and its conjugate cancel inside each same-tone covariance. In D1, the tap sums therefore yield common $C_{00,k}$ and $C_{10,k}$. In D2, with steering matrix $V$, tap-power matrix $P$, and motion-phase matrix $D$,
\begin{equation}
 C_{00,k}=VPV^H,\qquad C_{10,k}=VDPV^H.
\end{equation}
The invertible $V$ in D2 gives $A_k=VDV^{-1}$ for every tone. Thus the first two protocols do not create the proposed statistical dispersion at all.
D3 is different: its support-restricted operators have a maximum pairwise Frobenius distance of $0.8375$. The tested dispersion is real, but it produces no material learning advantage for the local method at the prescribed endpoint.

These tap models require checking dispersion of the forward--return statistical operator before paying for separate calibration.

\section{Spherical-Wavefront Focusing and Temporal Decorrelation}
The hypothesis that spherical-wavefront focusing should restore learning under temporal decorrelation was proposed in this study, motivated by coherent aperture gain in near-field beamforming~\cite{nearfield}. Classical distance-dependent propagation phase~\cite[Sec.~2.1]{tse} motivates compensating spherical path lengths.

Test E1 predicts near-field accuracy at least $95\%$ and a gain of at least $15$ pp over far-field correction. Accuracy below $85\%$ or gain below $8$ pp rejects it. E2, a LOS-plus-scattering model with Rician factor $3$, predicts at least $96\%$ accuracy and $20$ pp gain. Its corresponding rejection values are $88\%$ and $10$ pp. Both compare prescribed corrections at the same radial displacement.

The near-field advantage is $-0.2029$ pp in E1 and $-0.0092$ pp in E2 (Table~\ref{tab:e}). Both rejection rules trigger. The outcomes differ qualitatively: both corrections perform poorly in E1, while both perform well in E2. The paired accuracy intervals are $[-0.511,0.017]$ and $[-0.0214,0.0015]$ pp.

\begin{table}[t]
\caption{Test E: accuracy (\%) and near-field-minus-far-field gain (pp). The two rows use distinct channel models.}
\label{tab:e}\centering\footnotesize
\begin{tabularx}{\columnwidth}{@{}Xrrr@{}}\toprule
Model & Far field & Near field & Gain \\\midrule
\input{figures/table_e.tex} 
\bottomrule\end{tabularx}
\end{table}

E1 factors a single spherical spatial mode from a random scalar temporal gain. Aligning the spatial vectors cannot increase the magnitude of that scalar's correlation. At the tested displacement the scalar correlation is approximately $-0.000848$, so exact wavefront knowledge cannot supply the missing temporal information. The overlap between spherical and common radial phase compensation is already $0.992546$ in power. The large mismatch between initial plane and spherical steering vectors is a different quantity.

E2 adds a coherent LOS component whose cross-time outer product remains when the scattered component decorrelates. Both corrections retain useful LOS information. Its high near-field accuracy therefore does not show that curvature recovered the lost scattering coherence.

These single-mode and LOS-plus-white-scattering models separate spatial phase compensation from temporal predictability.

\section{Trace-Maximizing Port Assignment Under Nonsymmetric Motion}
The requirement of exact motion symmetry for successful port permutation was a working hypothesis proposed in this study.

Test F1 predicted accuracy below $65\%$ for every permutation at nonsymmetric directions, with a negative-gradient fraction of at least $50\%$. A prescribed candidate reaching $75\%$ rejects that claim. At displacement $0.5\lambda$ and direction $30^\circ$, a complete seven-port permutation reaches $99.5392\%$ accuracy (interval $[99.4781,99.6155]\%$), with $2.1875\%$ of minibatch updates having a negative real inner product in the return-noise-free diagnostic. This is the seed average of the fraction over $120$ updates comparing the estimated encoder gradient with the current true batch gradient. Forward noise and the effects of prior noisy training remain. F2's assignment-selected accuracies range from $99.5422\%$ to $99.6841\%$ on further nonsymmetric directions.

The permutations maximize the covariance-weighted score $\Rea\tr(C_{00}A^H\Pi)$. Their success rejects exact symmetry as a requirement.

\section{Zero-Mean Positioning Jitter Versus Fixed Displacement Bias}
The intuition that a jitter standard deviation and a fixed bias of the same magnitude should share a learning-collapse threshold was a working hypothesis in this study, prompted by the Bessel-function correlation zero in classical isotropic scattering~\cite[Sec.~2.4]{tse}. The archived protocols tested the competing prediction that the two error laws behave differently.

The archived protocols freeze the competing prediction that centered Gaussian jitter remains learnable while a comparable fixed bias does not. G1 rejects this jitter prediction if accuracy falls below $75\%$ at standard deviation $0.383\lambda$ or $0.45\lambda$. Its full target includes at least $85\%$ at both values and a $20$ pp advantage over bias at the former. G2 rejects a rigid-jitter advantage over bias below $20$ pp and targets at least $30$ pp. These are tests of the competing prediction, not a preregistered test that jitter moves a nonzero-displacement root.

At the common error parameter $0.383\lambda$, G1 gives $50.3235\%$ accuracy for fixed bias and $98.4222\%$ for independent jitter. G2 gives $53.2944\%$ for bias and $98.4909\%$ for rigid jitter (Table~\ref{tab:h}). The competing predictions meet their numerical conditions. Bias is a deterministic displacement and jitter's parameter is a standard deviation.

\begin{table}[t]
\caption{Test G: accuracy at bias magnitude or jitter standard deviation $0.383\lambda$. G1 and G2 use independent seed sets.}
\label{tab:h}\centering\footnotesize
\begin{tabularx}{\columnwidth}{@{}Xrr@{}}\toprule
Error process & Accuracy (\%) & 95\% interval \\\midrule
\input{figures/table_h.tex} 
\bottomrule\end{tabularx}
\end{table}

Both archives use a one-dimensional centered Gaussian positioning error in an isotropic field. With $e\sim\mathcal N(0,\sigma^2)$ in wavelength units, the same-port average is
\begin{align}
 \E[J_0(2\pi e)]
 &=\frac{1}{2\pi}\int_0^{2\pi}
 e^{-2\pi^2\sigma^2\sin^2\theta}\,d\theta\nonumber\\
 &=e^{-\pi^2\sigma^2}I_0(\pi^2\sigma^2)>0.
 \label{eq:jitter}
\end{align}
Averaging over centered errors is different from evaluating $J_0$ at a fixed bias. The positive expectation decays toward zero as jitter grows, and it does not ensure that every exchange is positively correlated. G2 records probability $29.6163\%$ of a negative scalar $J_0(2\pi e)$ at $\sigma=0.383\lambda$.

Equation~\eqref{eq:jitter} varies rail-jitter strength around zero residual displacement. For an independent isotropic vector error $\varepsilon\sim\mathcal N(0,\sigma^2I)$, the Gaussian characteristic function gives the separate identity
\begin{equation}
 \E_\varepsilon\phi(x+\varepsilon)=e^{-2\pi^2\sigma^2}\phi(x),
 \label{eq:isojitter}
\end{equation}
where $\phi(x)=\int e^{2\pi i u^Tx}\,\mu(du)$ and $\norm{u}=1$. The multiplier follows by taking the Gaussian characteristic function at each angular direction. It is positive, so averaging preserves existing real-correlation zeros and signs.

The rail-error protocols distinguish bias from jitter, while the isotropic-vector identity rejects zero removal at the correlation level.

\section{Scope}
The empirical conclusions apply to the small synthetic binary classifiers, channel laws, correction rules and finite training budgets specified in Section~I and Appendix~\ref{app:methods}. The quadratic recursion applies to its stated task subspace and noise model. Larger datasets, deeper networks, regression tasks and hardware deployments lie outside the tested domain. The interference comparisons use prescribed linear processing and independent messages. The wideband models use frequency-domain channels, and the near-field comparisons use single-mode or LOS-plus-white-scattering laws. The adjoint return is a baseband convention, not a validated conjugate transceiver protocol. The wideband experiments exclude intrasymbol Doppler and intercarrier interference. The frequency-dependent array case does not establish the physical sufficiency of its inherited nominal cyclic prefix. Empirical claims concern the recorded terminal comparisons and rejection thresholds. Causal identification and stationary limits are outside those measurements, while the exact same-setting neural-network forecast in Test A is outside the executed configurations. The training jitter results concern Gaussian rail errors. The isotropic-vector identity is an analytic result for a different error law.

Code and archived data are available from the corresponding author on reasonable request.

\section{Conclusion}
The seven cases identify where a physical-layer design rule needs a link to the training update distribution. In the quadratic proxy that link is the task-subspace second-moment recursion. The remaining comparisons test concrete performance forecasts under the stated tasks and budgets. A useful design prediction must specify task support, noise injection and the relevant update moments before inferring a training advantage from local phase, norm preservation or spatial alignment. Predictions fixed before evaluation and retained alongside their counterexamples make those inferences open to inspection.

\appendices
\input{methods.tex}

\bibliographystyle{IEEEtran}
\bibliography{refs}
\end{document}

%% file: figures/table_a.tex
Retained modes + innovation & Polar & 0.506785 & 3.262474 \\
Retained modes + innovation & TSVD & 0.502829 & 3.258438 \\
All modes + innovation & Polar & 1.025053 & --- \\
All modes + innovation & TSVD & 1.000000 & --- \\

%% file: figures/table_c.tex
C1: two users & 90.079 & [89.233, 90.921] & $\geq 75$ \\
C2: close user angles & 88.008 & [86.319, 89.548] & $\geq 80$ \\
C3: three users & 91.336 & [90.640, 92.081] & $\geq 82$ \\
C4: one self-interferer & 90.280 & [88.768, 91.626] & $\geq 75$ \\
C5: one AP alone & 90.646 & [90.302, 91.069] & $\geq 82$ \\
C6: two self-interferers & 93.181 & [92.703, 93.674] & $\geq 88$ \\
C7: AP phase jitter & 88.155 & [87.280, 89.249] & $\geq 82$ \\

%% file: figures/table_d.tex
D1: common tap statistics & 96.9160 & 96.9145 & -0.001526 \\
D2: distinct tap angles & 95.8996 & 95.8973 & -0.002289 \\
D3: frequency-dependent manifold & 97.5351 & 97.5433 & +0.008202 \\

%% file: figures/table_e.tex
E1: single spatial mode & 56.5598 & 56.3568 & -0.2029 \\
E2: LOS plus scattering & 99.7955 & 99.7864 & -0.0092 \\

%% file: figures/table_h.tex
G1: fixed bias & 50.3235 & [47.1573, 53.4624] \\
G1: independent jitter & 98.4222 & [98.2681, 98.5779] \\
G2: fixed bias & 53.2944 & [50.2869, 57.9941] \\
G2: rigid jitter & 98.4909 & [98.3658, 98.6221] \\
G2: independent jitter & 98.4711 & [98.3688, 98.5886] \\

%% file: methods.tex
\section{Operators and Frozen Experimental Configurations}
\label{app:methods}
This appendix specifies the configurations used for the comparisons reported in the text. Table~\ref{tab:tasks} supplies the learning budgets and noise levels. Distances below are in wavelengths. $\mathcal{CN}(0,C)$ denotes proper complex noise with covariance $C$, so a quoted standard deviation $s$ means complex variance $s^2$. Independent examples redraw the channel. A forward--return pair shares its specified latent field. Calibration, training, and test draws are separate, and methods within a seed share their underlying random draws. Evaluation root seeds and the two Philox master entropies are supplied in the accompanying source file \nolinkurl{supplement/seeds.json}. Classifier intervals use paired seed resampling with $20000$ percentile-bootstrap draws ($10000$ for C10). B uses $20000$ draws, and the quadratic simulation uses the $t$ interval specified below. All intervals are two-sided $95\%$ intervals.

\subsection{Signal conventions, learning, and calibration}
For matrix-channel tests the forward signal is $y=H_fWx+n_f$ and the returned derivative is $H_b^HM\delta+n_b$. This is an adjoint baseband convention, with conjugation included in the assumed return processing. Reciprocity alone gives a transpose and does not specify that processing. In the single-receiver interference tests, the forward signal is $y=h_0^HWx+n_f$, and the array receives $r=h_1s/|s|+z$ before applying a receive-side map $T$. Thus their correction acts after reception, whereas $M$ acts before the matrix-channel return. Zero derivative seeds transmit zero.

For either topology, let $u=\tanh(|y|^2)$, $p=\operatorname{sigmoid}(b+w^Tu)$, and
\begin{equation}
 \delta=2(p-\ell)\,w\odot(1-u^2)\odot y.
\end{equation}
The scalar version is denoted $s$. The encoder batch gradient is the mean of $g_xx^T$. Receiver gradients are computed locally, with $0.001w$ added to the weight gradient. Initially $W$ is circular Gaussian rescaled to Frobenius norm $0.4$, every receiver weight is $0.8$, and $b=-0.4$. Adam uses bias correction and $|g|^2$ for complex second moments. The bounds and rates in Table~\ref{tab:tasks} are applied after each update. Samples are consumed once in successive batches, including the final partial batch where present. Evaluation uses the final model on the initial-position forward channel, with the stated forward noise.

Except where specified otherwise below, statistical corrections use $64$ independently drawn noisy forward--return pilot pairs, each with proper-complex noise variance $0.01$ per entry. With $N_t$ independent transmit columns, the empirical moments are
\begin{align}
 \widehat C_{00}^{\rm raw}
 &=\frac{1}{64N_t}\sum_{j=1}^{64}\widehat F_j\widehat F_j^H-0.01I,\nonumber\\
 \widehat C_{10}
 &=\frac{1}{64N_t}\sum_{j=1}^{64}\widehat B_j\widehat F_j^H.
 \label{eq:pilotmoments}
\end{align}
The forward observation is quantized before these products. Hermitianize $\widehat C_{00}^{\rm raw}$, add $0.02I$, and floor its eigenvalues at $0.02$ to obtain $C_{\rm reg}$. Then $\widehat A=\widehat C_{10}C_{\rm reg}^{-1}$. For single-receiver tests $N_t=1$ in this estimator, and the subtracted pilot-noise covariance is that of the specified mixture and spatial filter, as detailed below. No correction is fitted to training outcomes.

Ordinary control quantization acts separately on real and imaginary parts: clip to $[-B,B]$, then round to the nearest of $2^{16}$ equally spaced levels including the endpoints, with ties to even. Use $B=4$ for forward pilot observations and $B=32$ for correction matrices. Quantized unitary matrices are projected back to the nearest unitary matrix by SVD. In A's SGD calibration and matrix interface, and in F2's polar reference, the signed quantizer instead uses the zero-preserving rule $B\,\operatorname{rint}(32767\,\operatorname{clip}(v/B,-1,1))/32767$. Unsigned norm feedback uses $65536$ equally spaced levels on its stated interval.

In A's SGD, D, E, and F1's polar reference, transmit $M\delta/\norm{M\delta}$ and restore the quantized norm after reception. Its public range is $[0,8\sqrt{N_r}\norm{M}_2]$. B's replay uses $[0,16]$, C1--C7 use $[0,8]$, and C8--C10 restore the exact $|s|$. F's assignment arms and G's bias/jitter arms instead transmit $M\delta/L$ and multiply the received vector by the fixed $L$ in Table~\ref{tab:tasks}. The resulting thermal-noise variance is $L^2\sigma_b^2$. These choices are part of the experiments, not interchangeable normalizations. CSI bypasses the return channel: D and G1 additionally quantize the exact digital encoder-input gradient at 16 bits with bound $32$. The other CSI references listed in the table use exact unquantized gradients.

\subsection{Test A: correction, task support, and distinct calibrations}
Order singular values decreasingly. If $A^H=U\Sigma V^H$, the two corrections are
\begin{align}
 M_{\rm polar}&=VU^H,\nonumber\\
 M_{\rm TSVD}&=V\operatorname{diag}
 \bigl(\sigma_i^{-1}\mathbf1[\sigma_i>\epsilon]\bigr)U^H.
 \label{eq:corrections}
\end{align}
The inverse gain is zero on a deleted mode. A complete SVD supplies the unitary extension for polar correction.

For the quadratic proxy, $p_i=i/2$, $i=0,1,2$, and $d=0.383$. Define the per-column spatial moments
\begin{align}
 C_{ij}&=J_0(2\pi(p_i-p_j)),\nonumber\\
 D_{ij}&=J_0(2\pi(p_i-p_j-d)),\nonumber\\
 A&=DC^{-1},\qquad Q=C-DC^{-1}D^H.
 \label{eq:a_geometry}
\end{align}
Use the exact matrices with absolute threshold $\epsilon=0.05$. No pilots, quantization, or transmit-seed normalization are applied. The retained task takes $U_r=U_{[:,1:2]}$, while the full task takes all three columns. Thus $H_f=U_r$ is deterministic. Each innovation column has covariance $Q/r$, rather than $Q$, and the control without innovation sets that covariance to zero. The initial state is $x_0=\mathbf1_r$. Step size, downstream noise, simulation length and burn-in are given in Section~I. Fresh circular Gaussian draws implement the conditional variance $x_k^HRx_k+0.03^2$. The paired simulation interval uses the eight post-burn-in seed means and a Student $t_7$ critical value.

The SGD comparison instead uses $d=0.38$, two independent transmit columns of the jointly Gaussian channel, and $H_f=C^{1/2}Z$, $H_b=AH_f+Q^{1/2}E$, with independent standard circular Gaussian $Z,E$. Its isotropic moments use $128$ equally spaced angular directions, equivalent to the Bessel kernel at the reported precision. Here the innovation covariance per column is $Q$, and both corrections are estimated from the same $64$ pilot pairs by~\eqref{eq:pilotmoments}. Apply~\eqref{eq:corrections} to $\widehat A^H$ with the relative cutoff $\epsilon=0.1\sigma_{\max}(\widehat A)$, quantize both matrices, and reproject only the polar matrix to be unitary. The downstream noise is added to the normalized physical return before norm recovery. This classifier uses the full random channel, without the proxy's retained-task restriction.

The method-specific SGD rate in Table~\ref{tab:tasks} was selected on three tuning seeds using mean terminal CE over $16$ geometries: isotropic $d\in\{0.38,0.40,0.42,0.55,0.70,0.609835\ldots\}$, and two angular spectra at $d\in\{0.1,0.2,0.3,0.4,0.5\}$. The last isotropic displacement is $j_{1,1}/(2\pi)$. The two spectra are a $0.7$ von Mises component with concentration $2$ plus $0.3$ uniform power, and a von Mises spectrum with concentration $50$, both centered at $35^\circ$. Discretize each on the same $128$ angles and normalize its weights. Selection minimizes the equal-weight mean across geometries and seeds, breaking ties toward the smaller rate. Both reported methods select $0.3$. Evaluation uses eight different seeds.

\subsection{Test B: field, selectors, and local statistic}
The channel-only comparison jointly samples positions $0$ and $0.383-j\Delta$, $j=0,\ldots,7$, with $\Delta=0.2,0.5$. Its separable covariance is $J_0(2\pi\norm{r-r'})C_t$, where $[C_t]_{tu}=J_0(\pi|t-u|)$, $t,u=0,\ldots,3$. A $256$-direction uniform angular quadrature constructs the joint covariance of all positions and the two analytic derivatives at the origin. Its Hermitian square root generates proper Gaussian samples. The kernel error relative to $J_0$ is below $10^{-10}$.

The forward port is always at zero. Max-power selection maximizes $\sum_t|B_{jt}|^2$. Random selection draws $j$ uniformly independently of the field. Fixed selection uses $j=0$. Geometric selection minimizes $|0.383-j\Delta|$, breaking ties toward the smaller index. Power scans are perfect and instantaneous. The statistic is
\begin{equation}
 c=\frac{\Rea\sum_t F_t^*B_{jt}}{\norm{F}\norm{B_j}},
 \qquad P(c>0).
\end{equation}
The local derivative uses only component $t=0$: $\partial_q\arg h=\Im(h^*\partial_qh)/|h|^2$, $q=x,y$. Amplitude bins have Rayleigh-decile boundaries $\sqrt{-\log(1-q)}$. The bright/dark ratio uses the empirical upper/lower $10\%$ within each seed. No phase unwrapping or clipping of dark samples is used. The curve in Fig.~\ref{fig:b} averages $1/|h|$ within each bin.

The separate training replay uses the same separable field with spacing $\Delta=1/7$ and max-power selection over its eight return ports. It reconstructs the original $7680$ training channels per seed. It does not retrain the models or substitute the channel-only spacings. Return processing has no phase or inverse correction.

\subsection{Tests C1--C7: propagation and calibrated receive chains}
For the single-path user models, put
\begin{align}
 a_m(\theta)&=e^{-i\pi m\cos\theta},\quad h_{j0}=z_ja(\theta_j),\nonumber\\
 h_{j1}&=\phi_jh_{j0},\quad
 \phi_j=e^{2\pi i(0.383)\cos\theta_j}.
\end{align}
Each $z_j\sim\mathcal{CN}(0,1)$ is independent across users and examples, and shared across the forward--return pair. C1 uses four array elements and angles $(30^\circ,60^\circ)$. C2 uses $(30^\circ,35^\circ)$. C3 uses $(20^\circ,45^\circ,70^\circ)$. The target is the first user. In C1/C2 the array return is $h_{11}t_1+\sqrt{10}h_{21}t_2+n_b$. C3 adds $\sqrt{10}h_{31}t_3$. Each $t_j=s_j/|s_j|$ is the normalized actual CE derivative from that user's independent XOR input and forward noise. Interfering models stay at independent initial parameters. Only the target learns, and forward resources are separate.

For the naive maps reported in C1--C4, form $\widehat A$ from mixed pilots and use the inverse unitary polar factor $Q_c=VU^H$ of $\widehat A=U\Sigma V^H$, quantized and reprojected as above. The output is $g_x=Q_cr\,q(|s_1|)$. C1/C2 mix the separately noisy pilots as $c_{1t}+\sqrt{10}c_{2t}$, giving noise covariance $0.11I$. C3 adds another such pilot, giving $0.21I$. These are pilot-noise covariances, distinct from data thermal variance $0.0009$.

C4 retains the $30^\circ$ target and places array receivers at $x_m=(-0.75,-0.25,0.25,0.75)$. A self-interference source at $(\sin15^\circ,\cos15^\circ)$ has signature $b_m=e^{-2\pi i r_m}/r_m$, rescaled to $\norm b^2=4$. At the reported $20$ dB point, the received contamination is $10bq$, where $q$ is independent equiprobable QPSK. Separate $64$-pair target and self-interference calibrations have variance $0.01$ each. Their sum has variance $0.02I$. The self-interference pilots use the same known QPSK symbol within a time pair. Their noise is not multiplied by the interference power.

C6 uses the same target and receiver locations, with two sources at $(\sin15^\circ,\cos15^\circ)$ and $(\sin(-35^\circ),\cos(-35^\circ))$. Normalize both spherical signatures to squared norm $4$ and set $H=\sqrt{100/2}[b_1,b_2]$. The contamination is $Hq$ with independent QPSK coordinates. In $64$ calibration pairs, the columns $q_k$ and $(-1)^kq_k$ are orthogonal. Then $\widehat H=\sum_k v_{1k}q_k^H/64$. Form
\begin{equation}
 P=I-\widehat H(\widehat H^H\widehat H)^{-1}\widehat H^H.
\end{equation}
Project both mixed pilot observations by $P$, subtract pilot-noise covariance $0.02P$, regularize, and obtain $Q_c$. The reported projected map is $T=PQ_cP$. It contains neither a target-gain inverse nor a target-direction reconstruction. The two sources split a total power ratio of $100$.

C5 has four equally strong, independent APs with two elements each and angles $(30,60,90,120)^\circ$. Apply the same $h_{l0},h_{l1}$ law to each two-element block. The full forward signal is $\sum_lh_{l0}^HW_lx+n_f$. Each AP's correction uses only its own $64$ noisy pilot pairs and a $2\times2$ estimate with noise subtraction $0.01I$. The reported single-AP method trains each AP in isolation, selecting within seed the largest terminal test accuracy, then smaller CE, then smaller AP index in a tie. This selection is explicitly a test oracle.

C7 uses that same AP geometry but draws, for every return example and independently for every calibration pair, $\vartheta_l\sim U[-\pi/3,\pi/3]$ independently across APs. Both antennas in one AP share the phase. The return is $e^{i\vartheta_l}(h_{l1}t+n_l)$. The old forward observation is unchanged. Centralized correction receives both old and new raw pilot vectors through a 4-bit real/imaginary quantizer with $16$ endpoint-inclusive levels on $[-4,4]$, before the ordinary 16-bit forward-pilot interface and the $8\times8$ estimate. Subtract $0.01I$, with no extra quantization-noise subtraction. Each data return vector is also 4-bit quantized before $Q_c$ and the 16-bit target norm recovery. No instantaneous AP phase is supplied to the estimator.

\subsection{Tests C8--C10: GSNR and processing order}
C8/C9 use the four-element, three-angle geometry of C3 with $d=0.383$. C8 replaces interfering model messages by fresh independent $u_2,u_3\sim\mathcal{CN}(0,1)$ and uses
\begin{equation}
 z=\sqrt{\rho_I/2}(a_2u_2+a_3u_3)+n_b,
 \quad \rho_I\in\{1,10,100\}.
\end{equation}
Its white-noise-only arms instead have covariance $10^{-s_b/10}I$, $s_b\in\{0,10,20\}$ dB, with no additional thermal term. Corrections are held fixed across data-noise powers. Calibration retains nominal power $10$ per interfering user. The main ranking has $13$ cells: three white/naive arms, nine multiuser arms (three powers times naive/ZF/MMSE), and CSI.

For these calibrated chains, obtain a spatial signature from the largest eigenpair $(\lambda,u)$ of a user's $64$-pilot covariance: rotate unit $u$ to make its first entry real nonnegative, then set $\widehat h=u\sqrt{\max(\lambda-0.01,0.02)}$. Naive uses $T=Q_c$. ZF obtains $P$ from the two estimated return signatures, projects both mixed pilot vectors, subtracts $0.21P$, fits $Q_c$, and uses $T=PQ_cP$. The loaded combiner uses
\begin{align}
 H_I&=\sqrt{10}[\widehat h_{21},\widehat h_{31}],\nonumber\\
 v&=(H_IH_I^H+0.1009I)^{-1}\widehat h_{11},\nonumber\\
 T_{\rm MMSE}&=\widehat h_{10}\,q_c\,v^H.
 \label{eq:loaded_chain}
\end{align}
Here $v^H$ first combines the mixed return. The factor $q_c$ is the scalar inverse polar phase fitted from the same mixed pilots after combining by $v^H$, and $\widehat h_{10}$ reconstructs the old target direction. In that scalar fit, quantize the old pilot at bound $4$, subtract $0.21\norm v^2$, add $0.02$, and floor at $0.02$. Quantize $q_c$ at bound $32$ and reproject it to the unit circle. The amplitude of $v$ and of the reconstructed signature is retained, without gain normalization.

For C8's GSNR, fix the common initial model and $8192$ common probe examples (the archived test-world draws). Let $G_i$ be the true single-example encoder parameter gradient, $B=\phi_1T$, $v_i=\operatorname{vec}(BG_i)$, $\mu=\overline v$, $\mu_0=\overline{\operatorname{vec}(G_i)}$, and $R_z=\E[zz^H]$. The trace variance of a batch of $32$ drawn with replacement from this empirical distribution is
\begin{align}
 V_g=\frac{1}{32}\bigl(&\overline{\norm{v_i}^2}-\norm\mu^2\nonumber\\
 &+\overline{|s_i|^2\norm{x_i}^2}\tr(TR_zT^H)\bigr).
\end{align}
The two reported metrics are
\begin{equation}
 \mathrm{GSNR}_{\rm vector}=\frac{\norm\mu^2}{V_g},\qquad
 \mathrm{GSNR}_{\rm scalar}=\frac{|\rho|^2\norm{\mu_0}^2}{V_g},
\end{equation}
where $\rho=a_1^HBa_1/4$. If $e=Ba_1-\rho a_1$, then $a_1^He=0$ and $\mathrm{GSNR}_{\rm vector}=(|\rho|^2+\norm e^2/4)/(V_g/\norm{\mu_0}^2)$. The scalar metric omits orthogonal mean energy. CSI includes minibatch sampling variance. The fixed-batch invariance diagnostic repeats the same $32$ examples with $4096$ independent return-noise draws. Rankings correlate the $13$ seed-mean GSNRs with terminal accuracies.

C9 fixes total data SIR at $0$ dB, namely $\rho_I=1$, while retaining C8's calibration power, maps, and exact norm recovery. Both SGD and Adam run single-user, ZF, and loaded-combiner arms. CSI uses Adam. The single-user control removes additive multiuser interference and fits its correction from target-only pilots with noise covariance $0.01I$. The loss ratio compares the two optimizers' single-minus-ZF seed-mean accuracy differences, at the common rates and budgets in Table~\ref{tab:tasks}.

C10 instead uses a $30^\circ$ target and one interferer at $45^\circ$, $60^\circ$, or $90^\circ$, with actual frozen-model CE messages and equal return powers. Spatial signatures and temporal phases are exact. There is no calibration or quantization. With $H=[\phi_1a_1,\phi_2a_2]$ and $f_1$ the first column of $H(H^HH+0.0009I)^{-1}$, the maps are
\begin{align}
 T_{\rm naive}=T_{\rm single}&=\phi_1^*I,\nonumber\\
 T_{\rm ZF}&=\phi_1^*\bigl(I-a_2a_2^H/4\bigr),\nonumber\\
 T_{\rm MMSE}&=a_1f_1^H.
\end{align}
The last map combines to a scalar and reconstructs the target direction. ZF has no such reconstruction or gain inverse. Single-user removes the interferer, and CSI uses the exact instantaneous gradient. At steps $80$ and $120$, use $256$ independent probe examples and $128$ independent interference/noise draws with their sign-reversed antithetic partners. Compute the cosine of the averaged encoder gradient with the true probe-batch gradient, then average the two checkpoints. The reported correlation uses $15$ angle--method seed means and merges theoretical ties within $10^{-10}$.

\subsection{Tests D1--D3: wideband channels and averaging order}
All three tests use eight tones, two independent transmit columns, $d=0.383$, and the per-tone learning task of Table~\ref{tab:tasks}. A whole wideband example shares its tap coefficients across tones. With normalized tap powers summing to one,
\begin{equation}
 H_t[k]=\sum_l\sqrt{p_l}\,h_{tl}e^{-2\pi i k\tau_l/8}.
\end{equation}
D1 has four taps with $\tau_l=l$, $p_l\propto e^{-l/2}$, $l=0,1,2,3$. Each column of $(h_{0l},h_{1l})$ is independently drawn with block covariance $\left[\begin{smallmatrix}C&D^H\\D&C\end{smallmatrix}\right]$ from~\eqref{eq:a_geometry}, using $128$ uniform directions and $d=0.383$. It uses three half-wavelength receiver elements and nominal CP length $4$.

D2 has three taps, angles $(30,60,90)^\circ$, delays $(0,1,2)$, and powers $(0.5,0.3,0.2)$. For $m=0,1,2$, set $a_m(\theta)=e^{-i\pi m\cos\theta}$ and draw independent $z_l\sim\mathcal{CN}(0,1)$ per example and transmit column. Then $h_{0l}=a(\theta_l)z_l$ and $h_{1l}=e^{2\pi i d\cos\theta_l}h_{0l}$. The nominal CP is $3$.

D3 retains these tap angles, delays, and powers, but uses four receiver elements and frequency ratios $r_k=1+(k-3.5)\eta/8$. The reported high-bandwidth comparison fixes $\eta=0.20$. Its array convention is explicitly $a_{km}(\theta)=e^{+i\pi m r_k\cos\theta}$, and its motion phase is $e^{2\pi i d r_k\cos\theta}$. Thus both array and motion phases vary with tone. The analytic $A_k=C_{10,k}C_{00,k}^{\dagger}$ uses relative pseudoinverse tolerance $10^{-10}$ on its rank-three support. Training uses full SVD unitary extensions of the regularized pilot estimates.

At each of $64$ pilot times, all tones and both transmit columns are observed. For the global method, average the eight raw $\widehat C_{00,k}$ and eight $\widehat C_{10,k}$ separately, then regularize, form $\widehat A$, and take its polar factor. For the local method, perform those steps independently per tone. Neither method averages polar factors. Both quantize each resulting matrix and reproject it to be unitary. Each tone has separate trainable parameters with a common within-seed initialization. Average the eight terminal tone metrics before seed-level inference.

\subsection{Tests E1/E2: spherical geometry and constrained fits}
Use $N_r=8$, $N_t=2$, $x_n=(n-3.5)/2$, $r_0=3$, $r_1=3.383$, and $\theta=60^\circ$ measured from the array normal. Define
\begin{align}
 d_n(r)&=\sqrt{r^2+x_n^2-2rx_n\sin\theta},\nonumber\\
 a_t[n]&=e^{-2\pi i d_n(r_t)}/\sqrt8,\nonumber\\
 a_{\rm FF}[n]&=e^{2\pi i x_n\sin\theta}/\sqrt8.
\end{align}
There is no $1/r$ path-loss factor. In E1, $u_t=\sqrt8a_t$, $F=u_0h_0$, $B=u_1h_1$, with $h_0,z\sim\mathcal{CN}(0,I_2)$ independent and $h_1=\rho h_0+\sqrt{1-|\rho|^2}z$. Here $\rho$ is the $128$-angle mean of $e^{-2\pi i(0.383)\cos\varphi}$, agreeing with $J_0(2\pi\,0.383)$. For near-field correction take $D=\operatorname{diag}(a_1/a_0)$. For far-field correction take $D=e^{-2\pi i(0.383)}I$. From the same regularized pilot moments, fit $\widehat\gamma=\tr(D^H\widehat C_{10})/\tr(C_{\rm reg})$ and use $M=\operatorname{phase}(\widehat\gamma)D$, with phase one at zero, followed by matrix quantization and unitary reprojection. No inverse magnitude is applied.

E2 uses unit-norm LOS vectors without E1's $\sqrt8$ scaling. Per transmit column,
\begin{equation}
 F=\sqrt{3/4}\,a_0+\tfrac12 Z_0,\qquad
 B=\sqrt{3/4}\,a_1+\tfrac12 Z_1,
\end{equation}
where $Z_0,Z\sim\mathcal{CN}(0,I_8/8)$ and $Z_1=\rho Z_0+\sqrt{1-|\rho|^2}Z$. Columns have independent scattering and the same LOS mean. This gives Rician factor $3$ and unit expected energy per column. Preserve the noncentered moments in calibration. Near-field correction uses the diagonal $D$ and fit above. Far-field correction uses $T_{\rm FF}=e^{-2\pi i(0.383)}a_{\rm FF}a_{\rm FF}^H$ and fits
\begin{equation}
 \widehat\gamma_{\rm FF}
 =\frac{\tr(T_{\rm FF}^H\widehat C_{10})}
 {\tr(T_{\rm FF}C_{\rm reg}T_{\rm FF}^H)}.
\end{equation}
Set $M=\operatorname{phase}(\widehat\gamma_{\rm FF})T_{\rm FF}$. After quantization, project back to the complex line spanned by $T_{\rm FF}$ and normalize its scalar coefficient to unit magnitude, preserving rank one. Thus E2's far-field arm is a projection, unlike E1's full-space phase operator. Both E models use the same $64$ pilot pairs for their compared fits.

\subsection{Tests F1/F2: finite-array assignment}
The seven-element array consists of the origin and $p_{k+1}=\tfrac12(\cos(k\pi/3),\sin(k\pi/3))$, $k=0,\ldots,5$. F1's reported nonsymmetric case has $d=0.5$ and direction $30^\circ$. F2 uses $d=0.383$, directions $15,37,73^\circ$, and both that array and the nine-element square $p=(x/2,y/2)$, $x,y=0,1,2$, in $x$-major order.

With $128$ uniform unit directions $u_l$, $v=(\cos\theta,\sin\theta)$, define $S_{0,il}=e^{-2\pi ip_i^Tu_l}/\sqrt{128}$ and $S_{1,il}=S_{0,il}e^{2\pi idv^Tu_l}$. Then $C_{00}=S_0S_0^H$, $C_{10}=S_1S_0^H$, $A=C_{10}C_{00}^{-1}$ and $Q=C_{00}-AC_{00}A^H$. Generate two independent columns with $F=C_{00}^{1/2}Z$, $B=AF+Q^{1/2}E$. Choose the complete permutation $\Pi_{\pi(i),i}=1$ maximizing $\Rea\tr(C_{10}^H\Pi)$. The score is normalized by the array size $7$ before ties are declared, so normalized scores within $10^{-12}$ (unnormalized within $7\times10^{-12}$) tie, resolved lexicographically. F1 enumerates all $7!$ permutations. F2 uses linear assignment with the same tie rule. Only geometry and the prescribed angular spectrum enter selection. No calibration is used for these assignments. The return-noise-free diagnostic recomputes each batch's physical encoder gradient with the return thermal term removed at the actual current trained parameters, retaining the forward noise.

\subsection{Tests G1/G2: sample the error before the field}
The four-element half-wavelength ULA has two independent transmit columns and rails perpendicular to its axis. Terminal displacement is $0.50$, ideally compensated mechanically by $-0.50$. The residual at port $i$ is $r_i=b-e_i$. Fixed bias uses $r_i=0.383$. Centered jitter uses $b=0$ and Gaussian errors of standard deviation $\sigma$. G1 uses independent errors across ports. G2 compares that law to one scalar error shared by all ports. Errors are redrawn independently each exchange.

Let $D_{ij}=(i-j)/2$. Conditional on the sampled residuals,
\begin{align}
 [C_{00}]_{ij}&=J_0(2\pi|D_{ij}|),\nonumber\\
 [C_{10}]_{ij}&=J_0(2\pi\sqrt{D_{ij}^2+r_i^2}),\nonumber\\
 [C_{11}]_{ij}&=J_0(2\pi\sqrt{D_{ij}^2+(r_i-r_j)^2}).
\end{align}
Set $A=C_{10}C_{00}^{-1}$, $Q=C_{11}-AC_{00}A^H$, and draw $F=C_{00}^{1/2}Z_F$, $B=AF+Q^{1/2}Z_B$ with independent standard circular Gaussian matrices. Zero residual explicitly gives $B=F$. Only roundoff-scale negative eigenvalues of $Q$ are clipped. An eigenvalue below $-10^{-9}$ fails validation. Random residuals must precede this conditional Gaussian draw. Replacing them by the mean covariance changes the mixture distribution. Bias/jitter arms use $M=I$, fixed return scale $L=16$, and no pilots or phase correction. The reported table uses $\sigma=0.383$. G1's additional supported endpoint uses $\sigma=0.45$ with the same law.

%% file: main.bbl
\begin{thebibliography}{10}
\providecommand{\url}[1]{#1}
\csname url@samestyle\endcsname
\providecommand{\newblock}{\relax}
\providecommand{\bibinfo}[2]{#2}
\providecommand{\BIBentrySTDinterwordspacing}{\spaceskip=0pt\relax}
\providecommand{\BIBentryALTinterwordstretchfactor}{4}
\providecommand{\BIBentryALTinterwordspacing}{\spaceskip=\fontdimen2\font plus
\BIBentryALTinterwordstretchfactor\fontdimen3\font minus
  \fontdimen4\font\relax}
\providecommand{\BIBforeignlanguage}[2]{{%
\expandafter\ifx\csname l@#1\endcsname\relax
\typeout{** WARNING: IEEEtran.bst: No hyphenation pattern has been}%
\typeout{** loaded for the language `#1'. Using the pattern for}%
\typeout{** the default language instead.}%
\else
\language=\csname l@#1\endcsname
\fi
#2}}
\providecommand{\BIBdecl}{\relax}
\BIBdecl

\bibitem{jsac}
``Keep pace to learn: Movable antennas restore the backpropagated gradient
  under user mobility,'' arXiv:2610.23118, 2026.

\bibitem{wcl}
``Wait to learn: Lobes and dead zones in physical backpropagation over an aging
  single-antenna channel,'' arXiv:2610.23135, 2026.

\bibitem{tcom}
``Wait or rotate: Statistical repair of the backward pass in over-the-air split
  learning under channel aging,'' arXiv:2610.23137, 2026.

\bibitem{aging}
\BIBentryALTinterwordspacing
K.~T. Truong and R.~W. Heath, Jr., ``Effects of channel aging in massive {MIMO}
  systems,'' arXiv:1305.6151, version 3, 2013. [Online]. Available:
  \url{https://arxiv.org/abs/1305.6151}
\BIBentrySTDinterwordspacing

\bibitem{tse}
\BIBentryALTinterwordspacing
D.~Tse and P.~Viswanath, \emph{Fundamentals of Wireless Communication}.\hskip
  1em plus 0.5em minus 0.4em\relax Cambridge University Press, 2005, chapters
  2, 3, and 8. [Online]. Available:
  \url{https://web.stanford.edu/~dntse/wireless_book.html}
\BIBentrySTDinterwordspacing

\bibitem{higham}
\BIBentryALTinterwordspacing
N.~J. Higham, ``Computing the polar decomposition---with applications,''
  \emph{SIAM Journal on Scientific and Statistical Computing}, vol.~7, no.~4,
  pp. 1160--1174, 1986. [Online]. Available:
  \url{https://eprints.maths.manchester.ac.uk/694/1/high86p.pdf}
\BIBentrySTDinterwordspacing

\bibitem{dieuleveut}
\BIBentryALTinterwordspacing
A.~Dieuleveut, A.~Durmus, and F.~Bach, ``Bridging the gap between constant step
  size stochastic gradient descent and markov chains,'' arXiv:1707.06386,
  version 2, 2018. [Online]. Available:
  \url{https://arxiv.org/abs/1707.06386v2}
\BIBentrySTDinterwordspacing

\bibitem{nyeberry}
\BIBentryALTinterwordspacing
J.~F. Nye and M.~V. Berry, ``Dislocations in wave trains,'' \emph{Proceedings
  of the Royal Society of London. Series A}, vol. 336, pp. 165--190, 1974.
  [Online]. Available:
  \url{https://michaelberryphysics.wordpress.com/wp-content/uploads/2013/07/berry034.pdf}
\BIBentrySTDinterwordspacing

\bibitem{bottou}
\BIBentryALTinterwordspacing
L.~Bottou, F.~E. Curtis, and J.~Nocedal, ``Optimization methods for large-scale
  machine learning,'' arXiv:1606.04838, version 3, 2018. [Online]. Available:
  \url{https://arxiv.org/abs/1606.04838v3}
\BIBentrySTDinterwordspacing

\bibitem{adam}
\BIBentryALTinterwordspacing
D.~P. Kingma and J.~L. Ba, ``Adam: A method for stochastic optimization,'' in
  \emph{International Conference on Learning Representations}, 2015. [Online].
  Available: \url{https://arxiv.org/abs/1412.6980}
\BIBentrySTDinterwordspacing

\bibitem{nearfield}
\BIBentryALTinterwordspacing
E.~Bj{\"o}rnson, {\"O}.~T. Demir, and L.~Sanguinetti, ``A primer on near-field
  beamforming for arrays and reconfigurable intelligent surfaces,''
  arXiv:2110.06661, version 2, 2021. [Online]. Available:
  \url{https://arxiv.org/abs/2110.06661v2}
\BIBentrySTDinterwordspacing

\end{thebibliography}
